\documentclass[twocolumn,prl,superscriptaddress,showpacs,floatfix,nobalancelastpage,preprintnumbers,nofootinbib]{revtex4}

\usepackage{epsfig}
\usepackage{bm}
\usepackage{color}

\long\def\dump#1{}

\newcommand{\D}{{\rm d}}

\newcommand{\E}{{\rm e}}

\begin{document}

\title{Multiple Spectral Splits of Supernova Neutrinos}

\author{Basudeb~Dasgupta}
\affiliation{Max-Planck-Institut f\"ur Physik
(Werner-Heisenberg-Institut), F\"ohringer Ring 6, 80805 M\"unchen,
Germany}

\author{Amol~Dighe}
\affiliation{Tata Institute of Fundamental Research, Homi Bhabha Road,
Mumbai 400005, India}

\author{Georg~G.~Raffelt}
\affiliation{Max-Planck-Institut f\"ur Physik
(Werner-Heisenberg-Institut), F\"ohringer Ring 6, 80805 M\"unchen,
Germany}

\author{Alexei~Yu.~Smirnov}
\affiliation{Abdus Salam International Centre for Theoretical Physics,
Strada Costiera 11, 34014 Trieste, Italy}
\affiliation{Institute for Nuclear Research, Russian Academy of
  Sciences, 117312 Moscow, Russia}

\date{19 April 2009, revised 17 June 2009}

\preprint{MPP-2009-33}

%%%%%%%%%%%%%%%%%%%%%%%%%%%%%%%%%%%%%%%%%%%%%%%%%%%%%%%%%%%%%%%%%%%%%%
\begin{abstract}
Collective oscillations of supernova neutrinos swap the spectra
$f_{\nu_e}(E)$ and $f_{\bar\nu_e}(E)$ with those of another flavor in
certain energy intervals bounded by sharp spectral splits. This
phenomenon is far more general than previously appreciated: typically
one finds one or more swaps and accompanying splits in the $\nu$ and
$\bar\nu$ channels for both inverted and normal neutrino mass
hierarchies.  Depending on an instability condition, swaps develop
around spectral crossings (energies where $f_{\nu_e}=f_{\nu_x}$,
$f_{\bar\nu_e}=f_{\bar\nu_x}$ as well as $E \rightarrow \infty$ where
all fluxes vanish), and the widths of swaps are determined by the
spectra and fluxes. Wash-out by multi-angle decoherence varies across
the spectrum and splits can survive as sharp spectral features.
\end{abstract}
%%%%%%%%%%%%%%%%%%%%%%%%%%%%%%%%%%%%%%%%%%%%%%%%%%%%%%%%%%%%%%%%%%%%%%


\maketitle

%%%%%%%%%%%%%%%%%%%%%%%%%%%%%%%%%%%%%%%%%%%%%%%%%%%%%%%%%%%%%%%%%%%%%%
{\em Introduction.}---%
%%%%%%%%%%%%%%%%%%%%%%%%%%%%%%%%%%%%%%%%%%%%%%%%%%%%%%%%%%%%%%%%%%%%%%
The neutrino flux from a core-collapse supernova (SN) is a powerful
probe of particle physics and astrophysics~\cite{Dighe:2008dq}.  SN
neutrinos interact not only with the stellar medium, producing the
Mikheyev-Smirnov-Wolfenstein (MSW) flavor conversion, but also with
other neutrinos and antineutrinos. The latter interactions modify the
flavor evolution in a non-linear fashion and give rise to collective
forms of oscillation \cite{Pantaleone:1992eq, Sigl:1992fn,
  Kostelecky:1994dt, Pastor:2001iu, Wong:2002fa, Balantekin:2006tg}, a
subject of intense recent investigation~\cite{Pastor:2002we,
  Sawyer:2005jk, Duan:2005cp, Duan:2006an, Hannestad:2006nj,
  Raffelt:2007yz, EstebanPretel:2007ec, Duan:2007mv, Raffelt:2007cb,
  Raffelt:2007xt, Duan:2007fw, Duan:2007bt, Fogli:2007bk,
  Fogli:2008pt, Duan:2007sh, Dasgupta:2008cd, Dasgupta:2007ws,
  Duan:2008za, Dasgupta:2008my, EstebanPretel:2008ni, Dasgupta:2008cu,
  Gava:2008rp, Raffelt:2008hr, Dasgupta2009}.

The most important observational consequence of the
collective effects is an exchange of the $\nu_e$ ($\bar\nu_e$)
spectrum with the $\nu_x$ ($\bar\nu_x$)
spectrum in certain energy intervals. We call such a flavor exchange
a ``swap'', whereas ``splits'' are sharp boundary features at the
edges of each swap interval. Spectral splits may become observable
in the high-statistics neutrino signal from the next galactic~SN,
leading to valuable clues about the underlying
physics~\cite{Duan:2007bt, Dasgupta:2008cd, Dasgupta:2008my}.

The well-understood ``classic swap'' covers the entire $\bar\nu$
spectrum and that of $\nu$ above an energy fixed by the approximate
conservation of the $\nu_e$ deleptonization flux \cite{Raffelt:2007cb,
  Raffelt:2007xt, Duan:2007fw}. In this paper we show that spectral
swaps and concomitant splits are more ubiquitous than has been
appreciated in the past. One example is the puzzling low-energy
split in the $\bar\nu$ spectrum that was noted for the inverted
neutrino mass hierarchy~\cite{Fogli:2008pt, Fogli:2007bk}.
However, with flavor spectra typical for SN neutrinos one
should expect multiple splits in either hierarchy.

We focus on neutrino-neutrino interactions alone and study two-flavor
oscillations driven by the atmospheric mass difference and 1--3
mixing. As has been established before~\cite{EstebanPretel:2007ec},
the usual matter effect in the region of collective oscillations (up
to a few 100~km) can be accounted for by choosing a small (matter
suppressed) effective mixing angle which we take to be $\theta_{\rm
  eff} = 10^{-5}$. MSW conversions occur typically at larger
distances. Their effects then factorize and can be included
separately~\cite{Dasgupta:2007ws}.

\begin{figure}[t]
\includegraphics[width=.9\columnwidth]{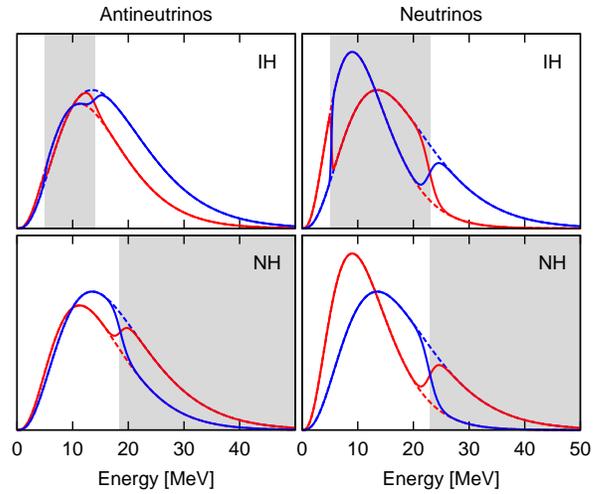}
\caption{SN neutrino spectra before (dashed lines) and after (solid
  lines) collective oscillations, but before possible MSW
  conversions. The panels are for $\nu$ and~$\bar\nu$, each time for
  IH and NH. Red lines $e$--flavor, blue $x$--flavor. Shaded regions
  mark swap intervals.\label{fig:Espectra}}
\end{figure}

%%%%%%%%%%%%%%%%%%%%%%%%%%%%%%%%%%%%%%%%%%%%%%%%%%%%%%%%%%%%%%%%%%%%%%
{\em Spectral crossings and spectral swaps.}---%
%%%%%%%%%%%%%%%%%%%%%%%%%%%%%%%%%%%%%%%%%%%%%%%%%%%%%%%%%%%%%%%%%%%%%%
Consider first the SN cooling phase where plausible choices
are~\cite{Raffelt:2003en}
$F_{\nu_e}:F_{\bar\nu_e}:F_{\nu_x}=0.85:0.75:1.00$ for the neutrino
fluxes, $\bar E_{\nu_e}=12$, $\bar E_{\bar\nu_e}=15$ and $\bar
E_{\nu_x}=\bar E_{\bar\nu_x}=18$~MeV for the average energies, and
$f_\nu(E)\propto E^3 e^{-4E/\bar E}$ for the spectral shape. Based on
the single-angle approximation for neutrino
propagation~\cite{Duan:2006an, EstebanPretel:2007ec, Fogli:2007bk,
  Dasgupta:2008cu}, Fig.~\ref{fig:Espectra} shows the flavor spectra
before and after collective oscillations. For the inverted mass
hierarchy (IH) we find a swap for both $\nu$ and $\bar\nu$ and thus
a total of four splits.  For the normal hierarchy (NH) the swaps
extend to infinite $E$, providing one split in the $\nu$~and
$\bar\nu$ spectrum each.

Flavor oscillations leave $f_{\nu_e}(E) + f_{\nu_x}(E)$ unchanged, so
$\Delta f_\nu(E)\equiv f_{\nu_e}(E)-f_{\nu_x}(E)$, and similarly for
$\bar\nu$, contains all relevant information.  It proves crucial for
understanding the multisplit phenomenon (i)~to use $-\Delta
f_{\bar\nu}(E)$ for $\bar\nu$ (the ``flavor isospin'' construction
\cite{Duan:2006an}), and (ii)~to merge the $\nu$ and $\bar\nu$ spectra
to a single continuum in terms of the variable $\omega\equiv\pm|\Delta
m^2/2E|$, the vacuum oscillation frequency, that is positive for $\nu$
and negative for $\bar\nu$. Here $\Delta m^2 \equiv m_2^2-m_1^2$ with
$\nu_1 \approx \nu_e$ and $\nu_2 \approx \nu_x$, so $\Delta m^2>0$
stands for NH and $\Delta m^2<0$ for IH. With $E_\omega =|\Delta m^2/2
\omega|$ we thus define our spectrum~as
\begin{equation}
\label{eq:gdef}
g_\omega \equiv \frac{|\Delta m^2|}{2\omega^2}
\times\cases{f_{\nu_e}(E_\omega)-f_{\nu_x}(E_\omega)&for $\omega>0$\cr
f_{\bar\nu_x}(E_\omega)-f_{\bar\nu_e}(E_\omega)&for $\omega<0$\cr}~,
\end{equation}
where the common factor comes from the $E\to\omega$ transformation.
In  SN  we have $f_{\nu_x}(E)=f_{\bar\nu_x}(E)$ and a net
$\nu_e$ deleptonization flux, so $\int_{-\infty}^{+\infty}\D\omega\,
g_\omega>0$. Notice that $g_\omega>0$ means excess $\nu_e$ for
$\omega>0$ and excess $\bar\nu_x$ for $\omega<0$.

\begin{figure}
\includegraphics[width=0.8\columnwidth]{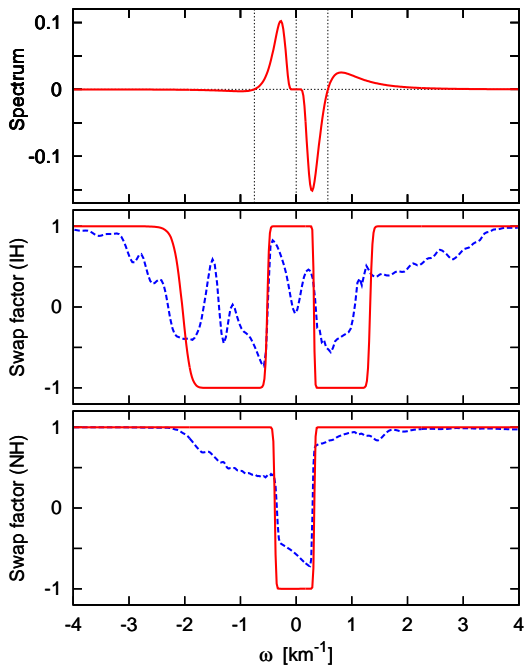}
\caption{Initial spectrum $g_\omega$ defined in
  Eq.~(\ref{eq:gdef}) for the cooling-phase example of
  Fig.~\ref{fig:Espectra}. Swap factor after collective oscillations
  for IH and NH as indicated. Single-angle (solid) and multi-angle
  (dashed). For animations see Ref.~\cite{movies}.
  \label{fig:omegaspectra}}
\end{figure}

\begin{figure}
\includegraphics[width=0.8\columnwidth]{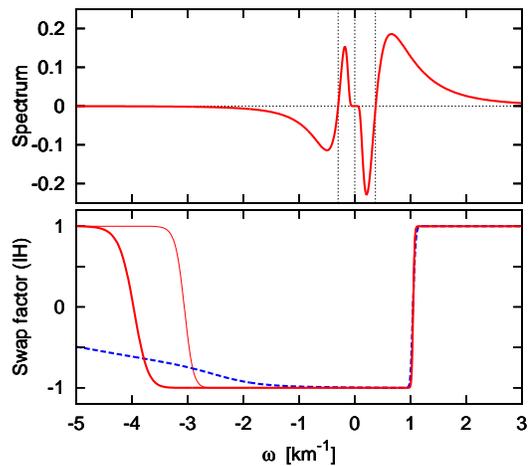}
\caption{Same as Fig.~\ref{fig:omegaspectra} for the accretion-phase
  example of Refs.~\cite{Fogli:2008pt,Fogli:2007bk}. Thin solid line:
  swap factor for $\theta_{\rm eff}=10^{-10}$ instead of
  $10^{-5}$. There is no resolved swap for
  NH.\label{fig:omegaspectra-italian}}
\end{figure}

In Fig.~\ref{fig:omegaspectra} we show $g(\omega)$ corresponding to
the example of Fig.~\ref{fig:Espectra}. The effect of collective
oscillations is represented by a ``swap factor'' $S_\omega$, giving
the final spectrum as
\begin{equation}\label{eq:swap}
 g_\omega^{\rm final} = S_\omega\,g_\omega^{\rm initial} \,.
\end{equation}
The ``spectral crossings'' where $g(\omega)=0$ are indicated by
vertical dotted lines. In particular, our construction provides a
$\nu$-$\bar{\nu}$ crossing at $\omega=0$ (infinite $E$), playing the
same role as a ``true flavor crossing'' where
$f_{\nu_e}(E)=f_{\nu_x}(E)$. The $\nu$-$\bar{\nu}$ crossing exists
even if the source emits only $\nu_e$ and~$\bar\nu_e$.

Figure~\ref{fig:omegaspectra} illustrates a general point: for IH
every $g_\omega$ crossing with a positive slope is an ``unstable
crossing'', i.e.\ it induces a swap over a finite $\omega$ range
with splits that become sharp in the adiabatic limit. For NH every
$g_\omega$ crossing with a negative slope is an unstable crossing.
(The equations of motion actually reveal that the dynamics for a
given spectrum $g_\omega$ in IH is equivalent to a spectrum
$-g_\omega$ in NH, thus there is no fundamental difference between
the two hierarchies.)

The usual assumptions about SN neutrinos lead to a triple-crossed
spectrum and thus to one or two swaps (two or four splits), depending
on the hierarchy. An apparent counter example, representing the SN
accretion phase, is $\bar E_{\nu_e}=10$, $\bar E_{\bar\nu_e}=15$ and
$\bar E_{\nu_x}=\bar E_{\bar\nu_x}=24$~MeV and
$F_{\nu_e}:F_{\bar\nu_e}:F_{\nu_x}=2.4:1.6:1.0$~\cite{Fogli:2008pt,
  Fogli:2007bk}. The corresponding $\omega$ spectrum
(Fig.~\ref{fig:omegaspectra-italian}) has three crossings, yet only
one swap is observed for IH and none for NH. Below we argue that the
narrow spacing of the crossings implies that (i)~in the adiabatic
limit the missing swap is exponentially narrow, and (ii)~it is
suppressed by adiabaticity violation. Therefore, a narrowly spaced
triple crossing can superficially act like a single one.

\begin{figure}
\includegraphics[width=0.67\columnwidth]{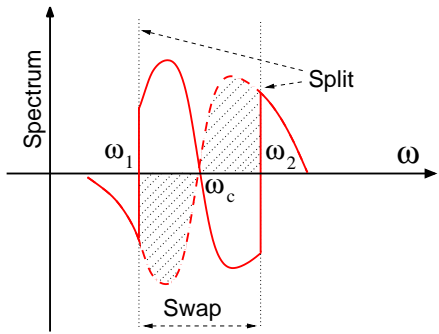}
\caption{Basic case of a spectrum producing a swap.
The initial spectrum (dashed) has an unstable crossing at
$\omega_c$, whereas the final spectrum (solid) has splits at
$\omega_1$ and $\omega_2$.
\label{fig:schematic}}
\end{figure}

%%%%%%%%%%%%%%%%%%%%%%%%%%%%%%%%%%%%%%%%%%%%%%%%%%%%%%%%%%%%%%%%%%%%%%
{\em Single-crossed system.}---%
%%%%%%%%%%%%%%%%%%%%%%%%%%%%%%%%%%%%%%%%%%%%%%%%%%%%%%%%%%%%%%%%%%%%%%
Many of our general observations are borne out from a generic
single-crossed example for IH (Fig.~\ref{fig:schematic}). The
equations of motion (EOMs) are
\begin{equation}\label{eq:EOM1}
\dot{\bf P}_{\omega}
=(\omega{\bf B}+\mu{\bf P})\times{\bf P}_{\omega}\,,
\end{equation}
where ${\bf B}$ is a unit vector $(0,0,1)$ in flavor space, ${\bf
  P}_{\omega}$ is the polarization-vector spectrum that covers $\nu$
for $\omega>0$ and $\bar\nu$ for $\omega<0$, and ${\bf
  P}=\int_{-\infty}^{+\infty}\D\omega\, {\bf P}_{\omega}$. The
coupling $\mu$ depends on $G_{\rm F}$, the local $\nu$ and $\bar\nu$
densities, and the normalization of $g_\omega$. Initially ${\bf
  P}_\omega=(0,0,g_\omega)$, for a vanishingly small mixing angle. The
crossing is unstable for IH because a swap would decrease the energy
$\int \D\omega\;\omega{\bf B}\cdot {\bf P}_{\omega}$. (For NH, the
$\omega {\bf B}$ term in Eq.~(\ref{eq:EOM1}) appears with a negative
sign, so $\int \D\omega \, \omega {\bf B} \cdot {\bf P}_\omega$ is
already minimal. However, a spectrum $-g_\omega$ would have an
unstable crossing.) We seek the final spectrum for this schematic
case, when $\mu$ slowly decreases from some initial value $\mu_0$
to~0.

The EOMs imply that ${\bf B}\cdot{\bf P}=\int\D\omega\,g_\omega$ is
conserved, so we can ``rotate away'' a common precession of the
entire system with frequency $\mu{\bf B}\cdot{\bf P}$ in the same
way as removing the usual matter term~\cite{EstebanPretel:2008ni}.
The EOMs become
\begin{equation}\label{eq:EOM2}
\dot{\bf P}_{\omega}
=(\omega\,{\bf B}+\mu\,{\bf P}_\perp)\times{\bf P}_{\omega}\,,
\end{equation}
where ${\bf P}_\perp$ is the component of ${\bf P}$ transverse to ${\bf
B}$.

For $ {\bf B} \cdot {\bf P} \not=0$ there is a critical value
$\mu_{\rm sync}$ such that for $\mu_0>\mu_{\rm sync}$ the system
initially performs synchronized oscillations with $\mu{\bf P}_\perp =
0$.  As $\mu$ decreases, each ${\bf P}_\omega$ follows its Hamiltonian
$\omega{\bf B}+\mu{\bf P}_\perp$, performing a ``pure precession,''
and all ${\bf P}_\omega$ with $\omega<\omega_{\rm split}$ end
anti-aligned with ${\bf B}$, the others aligned, leading to the
classic split fixed by ${\bf B}\cdot{\bf P}$ conservation
\cite{Raffelt:2007cb, Raffelt:2007xt, Duan:2007fw}. For ${\bf
  B}\cdot{\bf P}=0$ and $\mu_0\to\infty$, the system evolves like the
flavor pendulum~\cite{Hannestad:2006nj}, ending with the entire
spectrum swapped.

A qualitatively different behavior occurs for $\mu_0<\mu_{\rm sync}$,
or any finite $\mu_0$ if ${\bf B}\cdot{\bf P}=0$. Now $\mu{\bf
  P}_\perp$ performs a fast non-adiabatic motion with large amplitude,
causing a collective oscillation of the entire spectrum.  This generic
behavior is borne out analytically for a spectrum that is
antisymmetric in $\omega$ relative to $\omega_{\rm c}$, leading to a
``pure nutation'' in terms of a variable $\varphi(t)$.  Without loss
of generality we use $\omega_{\rm c}=0$ and find \cite{Raffelt:2008hr,
  Dasgupta2009},
\begin{equation}\label{eq:solution}
{\bf P}_{\omega}=
\frac{g_{\omega}}
{\sqrt{\omega^4-2\omega^2\kappa^2c_{\rm m}+\kappa^4}}
\pmatrix{-\kappa^2 s\cr
\omega\kappa\sqrt{2(c-c_{\rm m})}\cr
\omega^2-\kappa^2 c\cr}
\end{equation}
where $s\equiv\sin\varphi$, $c\equiv\cos\varphi$, $c_{\rm m}\equiv
\cos\varphi_{\rm max}$, and $\varphi(t)$ obeys
$\frac{1}{2}\dot{\varphi}^2= \kappa^2(\cos\varphi-\cos\varphi_{\rm
  max})$. Differentiation provides $\ddot{\varphi} =-
\kappa^2 \sin \varphi$, the equation for an anharmonic oscillator.
Equation~(\ref{eq:solution}) solves the EOM~(\ref{eq:EOM2})
if the nutation amplitude $\varphi_{\rm max}$ and frequency $\kappa$
fulfill the consistency relation
\begin{equation}\label{eq:consistency1}
\int\D\omega\,\frac{\omega\,g_{\omega}}
{\sqrt{\omega^4-2\omega^2\kappa^2c_{\rm m}+\kappa^4}}
=\frac{1}{\mu}\,.
\end{equation}
For a ``symmetric box'' where
$g(\omega)=1/\Delta\omega$ for $0<\omega<\frac{1}{2}\Delta\omega$
and $g_{-\omega} =-g_{\omega}$ we find
$\kappa_0=\Delta\omega/\sqrt{\E^{\Delta\omega/\mu_0}-1}$.

The initial state when $\mu=\mu_0$ is described by
$\varphi=\varphi_{\rm max}=\pi$, implying $c_{\rm m}=-1$ and so the
amplitude of ${\bf P}_\omega^z$ in Eq.~(\ref{eq:solution}) is
$\kappa_0^2/(\omega^2+\kappa_0^2)$. The modulation of $g_\omega$ thus
has a resonance with a maximum at $\omega = 0$.  As $\mu$ decreases
adiabatically from $\mu_0$ to 0, the r.h.s.\ of
Eq.~(\ref{eq:consistency1}) becomes singular and so the l.h.s.\ must
become singular as well. This is only possible for $c_{\rm m}\to +1$,
i.e., the final maximum nutation amplitude is $\varphi_{\rm
  max}=0$. At the same time, the frequency $\kappa$ decreases from
$\kappa_0$ to some non-zero value $\kappa_{\rm s}$. With $c_{\rm
  m}=+1$, Eq.~(\ref{eq:solution}) provides
\hbox{$S_\omega=(\omega^2-\kappa_{\rm s}^2)/|\omega^2-\kappa_{\rm
    s}^2|=\pm 1$}, and thus there is a swap in the range of
frequencies $\omega^2 < \kappa_{\rm s}^2$.
In other words, the final
$\kappa$ plays the role of the split frequency $\omega_{\rm
  split}=\pm\kappa_{\rm s}$, which  
is related to $\kappa_0$ through an adiabatic invariant.

Because $\kappa_0$ plays the dual role of the approximate swap width
and an inverse evolutionary time scale, an exponentially small
$\kappa_0$ implies that the corresponding swap disappears (i)~by not
being numerically resolved and (ii)~by adiabaticity violation.

A non-zero mixing ($0<\theta_{\rm eff} \ll 1$) is required to
trigger the initial motion, but also causes a delay until the system
relaxes to the adiabatic solution. Therefore, the effective $\mu_0$
relevant for swap formation is reduced and the swap width is smaller
for a smaller $\theta_{\rm eff}$. This subtle effect can become
visible when a split falls into a tail of $g_\omega$ where the flux
is small. A case in point is the left split in
Fig.~\ref{fig:omegaspectra-italian} that shifts to the right for a
smaller $\theta_{\rm eff}$ (thin-line), explaining the dependence on
the matter profile (that modifies $\theta_{\rm eff}$) noted in
Refs.~\cite{Fogli:2008pt, Fogli:2007bk}. Similar remarks
pertain to the left-most IH split in Fig.~\ref{fig:omegaspectra}.

%%%%%%%%%%%%%%%%%%%%%%%%%%%%%%%%%%%%%%%%%%%%%%%%%%%%%%%%%%%%%%%%%%%%%%
{\em Multiple crossings.}---%
%%%%%%%%%%%%%%%%%%%%%%%%%%%%%%%%%%%%%%%%%%%%%%%%%%%%%%%%%%%%%%%%%%%%%%
For a spectrum with several crossings the evolution is more
complicated and depends on various features, notably the separation
of the crossings and $\mu_0$. For instance, a multi-crossed system
can be constructed from two clones of Fig.~\ref{fig:schematic} with
a large $\omega$ range of empty modes between them. If $\mu_0$ is
small, each individual block shows the behavior discussed above.
Even though the entire spectrum feels the same $\mu{\bf
P}_\perp(t)$, only the Fourier components close to a given $\omega$
strongly affect ${\bf P}_\omega$, so different regions of the
spectrum can evolve almost independently of each other. If
$\mu_0>\mu_{\rm sync}$ of a single clone, we get the ``co-rotating
plane solution'' and classic split in each clone, once more because
distant Fourier components of ${\bf P}_\perp$ do not affect a given
solution. If $\mu_0$ is so large that both clones are synchronized
with each other, a true adiabatic solution would lead to one single
classic split for the entire spectrum, but this solution is
unstable. Once $\mu$ has become smaller than the separation between
the clones, the classic co-rotation solution for the entire spectrum
gives way to two approximately independent solutions, one for each
clone.

Most interesting in the SN context is a triple-crossed spectrum. For
NH we have a single unstable crossing flanked by two stable ones,
resulting in a single swap (Fig.~\ref{fig:omegaspectra}) that can be
understood, using the appropriate $g_\omega$ in
Eq.~(\ref{eq:consistency1}).  For schematic spectra one finds
explicitly an exponential decrease of $\kappa_0$ with decreasing
spacing of the crossings. For IH the crossings are
unstable-stable-unstable and generically provide two swaps. However,
they can merge to a single swap if the spacings between the crossings
are too small, i.e.\ the triple crossing acts as an effective single
one. The dynamical formation of the swaps in the triple-crossed case
can be very different depending on the spectrum.  The location and
widths of the swaps are fixed dynamically by coupled-pendulum
solutions together with the corresponding adiabatic invariants.

%%%%%%%%%%%%%%%%%%%%%%%%%%%%%%%%%%%%%%%%%%%%%%%%%%%%%%%%%%%%%%%%%%%%%%
{\em Multi-angle effects.}---%
%%%%%%%%%%%%%%%%%%%%%%%%%%%%%%%%%%%%%%%%%%%%%%%%%%%%%%%%%%%%%%%%%%%%%%
SN~neutrinos travel on different trajectories and every mode is
described by its direction of motion, in addition to~$\omega$.
Multi-angle effects can cause kinematical decoherence among angular
modes~\cite{Raffelt:2007yz} and smear the spectral features.
However, in spherically symmetric cases the collective motions seem
rather robust against multi-angle effects and a single-angle
treatment often seems to capture the main
effect~\cite{EstebanPretel:2007ec}.

The swap factors from multi-angle simulations of our examples are
shown as dashed lines in Figs.~\ref{fig:omegaspectra}
and~\ref{fig:omegaspectra-italian}. The right split in
Fig.~\ref{fig:omegaspectra-italian} is robust whereas the left one
smears out as previously noted~\cite{Fogli:2008pt, Fogli:2007bk}. In
Fig.~\ref{fig:omegaspectra} we find significant decoherence across
the spectrum. The IH case requires a large number of modes to reach
apparent convergence (we used 200 frequency and 15000 angular
modes). These issues require a dedicated investigation that is
beyond the scope of our work. It seems clear, however, that some of
the splits survive multi-angle decoherence as sharp spectral
features.

%%%%%%%%%%%%%%%%%%%%%%%%%%%%%%%%%%%%%%%%%%%%%%%%%%%%%%%%%%%%%%%%%%%%%%
{\em Concluding remarks.}---%
%%%%%%%%%%%%%%%%%%%%%%%%%%%%%%%%%%%%%%%%%%%%%%%%%%%%%%%%%%%%%%%%%%%%%%
Spectral swaps with concomitant splits are generic for dense neutrino
fluxes. They can appear in the $\nu$ and $\bar\nu$ channels for both
mass hierarchies. We have provided a qualitative explanation for the
main features of the multi-split phenomenon.  Spectral swaps are
nucleated by unstable crossings of the initial flavor spectra. The
number and locations of observationally relevant splits depends on the
properties of the initial spectra and the effective mixing angle,
opening the possibility for observing interesting time-dependent
features. Notice that the SN flavor spectra strongly differ between
the prompt deleptonization burst, the accretion phase, and the cooling
phase. Multi-angle effects smear some of the splits, once more opening
the possibility for interesting time-dependent features.  It seems
that the physics of SN neutrino conversion is much richer than it was
thought before and further studies may uncover a number of
qualitatively new phenomena.

%%%%%%%%%%%%%%%%%%%%%%%%%%%%%%%%%%%%%%%%%%%%%%%%%%%%%%%%%%%%%%%%%%%%%%
{\em Acknowledgements.}---%
%%%%%%%%%%%%%%%%%%%%%%%%%%%%%%%%%%%%%%%%%%%%%%%%%%%%%%%%%%%%%%%%%%%%%%
We thank A.~Mirizzi and E.~Lisi for fruitful discussions and
acknowledge partial support by the Deutsche Forschungsgemeinschaft
under grant TR-27 ``Neutrinos and Beyond'' and the Cluster of
Excellence ``Origin and Structure of the Universe'' (BD and GR), by
a Max Planck--India Partnergroup Grant (AD) and by the Alexander von
Humboldt-Foundation (AS).

%%%%%%%%%%%%%%%%%%%%%%%%%%%%%%%%%%%%%%%%%%%%%%%%%%%%%%%%%%%%%%%%%%%%%%
%%%  Bibliography  %%%%%%%%%%%%%%%%%%%%%%%%%%%%%%%%%%%%%%%%%%%%%%%%%%%
%%%%%%%%%%%%%%%%%%%%%%%%%%%%%%%%%%%%%%%%%%%%%%%%%%%%%%%%%%%%%%%%%%%%%%

%%%%%%%%%%%%%%%%%%%%%%%%%%%%%%%%%%%%%%%%%%%%%%%%%%%%%%%%%%%%%%%%%%%%%%
\end{document}